\documentclass[preprint]{imsart}

\RequirePackage[OT1]{fontenc}
\RequirePackage{amsthm,amsmath}
\RequirePackage{natbib}
\RequirePackage{hyperref}

\arxiv{arXiv:0000.0000}

\usepackage{multirow}
\usepackage{booktabs}
\usepackage{lscape}
\usepackage{mathtools}
\usepackage{float}
\usepackage{bibentry}
\usepackage{adjustbox}

\begin{document}

\begin{frontmatter}
\title{Multivariate Spatial-Temporal Variable Selection with Applications to Seasonal Tropical Cyclone Modeling}
\runtitle{Multivariate Spatial-Temporal Variable Selection}

\begin{aug}
\author{\fnms{} \snm{Marcela Alfaro C\'{o}rdoba}\thanksref{m1}\ead[label=e1]{malfaro@ncsu.edu}},
\author{\fnms{} \snm{ Montserrat Fuentes}\thanksref{m1}\ead[label=e2]{fuentes@ncsu.edu}},
\author{\fnms{} \snm{Joseph Guinness}\thanksref{m1}\ead[label=e3]{jsguinne@ncsu.edu}}
\and
\author{\fnms{} \snm{Lian Xie}\thanksref{m2}
\ead[label=e4]{xie@ncsu.edu}
\ead[label=u1,url]{http://www.cfdl.meas.ncsu.edu}}
\runauthor{Alfaro-C\'{o}rdoba et al.}

\address{Department of Statistics, North Carolina State University\thanksmark{m1} and Department of Marine, Earth and Atmospheric Sciences, North Carolina State University \thanksmark{m2}}


\end{aug}

\begin{abstract}
Tropical cyclone and sea surface temperature data have been used in several studies to forecast the total number of hurricanes in the Atlantic Basin. Sea surface temperature (SST) and latent heat flux (LHF) are correlated with tropical cyclone occurrences, but this correlation is known to vary with location and strength of the storm. The objective of this article is to identify features of SST and LHF that can explain the spatial-temporal variation of tropical cyclone counts, categorized by their strength. We develop a variable selection procedure for multivariate spatial-temporally varying coefficients, under a Poisson hurdle model (PHM) framework, which takes into account the zero inflated nature of the counts. The method differs from current spatial-temporal variable selection techniques by offering a dynamic variable selection procedure, that shares information between responses, locations, time and levels in the PHM context. The model is used to study the association between SST and LHF and the number of tropical cyclones of different strengths in 400 locations in the Atlantic Basin over the period of 1950-2013. Results show that it is possible to estimate the number of tropical storms by season and region. Furthermore, the model delimits areas with a significant correlation between SST and LHF features and the occurrence and strength of TCs in the North Atlantic Basin.
\end{abstract}

%

\end{frontmatter}

\section{Introduction}

A tropical cyclone is a rotating system characterized by a low-pressure center, strong winds, and heavy rain that can cause severe damage both out at sea and inland. Because of the numerous deaths and damages caused in the past (\cite{Xie2014b}), it is crucial to develop methods to model and predict the locations and strengths \footnote{Throughout this paper we use the word strength to refer to the classification of the storm according to its strength. We use the word intensity to refer to the $\lambda$ parameter in the Poisson hurdle model.} of storms. According to \cite{Keith2009}, a hurricane is an environmental heat engine driven by sensible and latent heating from the ocean. Measures of sea surface temperature (SST) and latent heat flux (LHF) values in certain areas and during some months of the year are correlated with cyclone activity (\cite{Gray1994}, \cite{Xie2014b} and \cite{Xie2005}), and thus should be important variables to describe the number of tropical cyclones (TC) in space and time. 

Numerous studies have described the relationship between SST and the number of TCs per season, with independent models per subregions in the North Atlantic Basin, delimited by $10^{\circ}N$ to $62^{\circ}N$ latitude and $10^{\circ}W$ to $110^{\circ}W$ longitude; see Figure~\ref{fig:mapall}. \cite{Lehmiller1997}, \cite{Blake2004}, \cite{Xie2005}, \cite{Elsner2006},  \cite{Keith2009}, \cite{Werner2011} and \cite{Xie2014} use climate indices as covariates in models that predict the number of hurricanes in each of the following subregions: Caribbean Ocean, Gulf of Mexico and the rest of the North Atlantic Basin. While some of the climate indices are summaries of SST from specific geographical locations and months of the year, none of the models in the literature study all possible combinations of months and locations. 

Previous investigations using spatial models have determined that it is possible to describe the spatial variation of hurricanes using a finer scale than the regional analyses discussed above. \cite{Elsner2000} show that the location of hurricanes on a grid can be modeled using spatial models and climate indices. \cite{Jagger2002} develop a space time autoregressive model and \cite{Hodges2014} use a Bayesian hierarchical spatial model, with total hurricanes by location as a response and climate indices per location as covariates. \cite{Jagger2002} investigate forecasting skill for the model when it predicts hurricane activity for each year in each location, but pointed out that their estimators failed to converge using a grid size smaller than 5 by 5 degrees. 

One important feature of the TC data on a grid is that the number of storms in a given year is likely to be zero in most of the locations in the Atlantic. Since the probability of an occurrence is directly related to the expected number of occurrences in a year, we propose a model that appropriately model both parameters and its association. \cite{Neelon2013} propose the Poisson Hurdle Model (PHM) for such cases, as they warn that failing to account for this dependence in zero-inflated models may produce biased parameter estimates. 

In this paper, we model the spatial-temporal variation of the yearly number of three strengths of TCs in the Atlantic Basin, and estimate their respective multivariate spatial-temporal association with SST and LHF. Simultaneously modeling the multivariate spatial-temporal process improves our ability to find features in the space-time covariates that are significantly associated with TC activity in specific locations. To this end, we propose a model to address the following challenges: (1) a multivariate response that is potentially zero inflated, (2) an association between response and covariates that varies over space and time, (3) presence of correlation among coefficients for different responses, locations and times. We use empirical orthogonal functions (EOFs) to create fields of SST and LHF features, and use their scores as covariates. We show that the multivariate spatial-temporal variation of TCs can be explained by spatial-temporal features of SST and LHF, using high spatial resolution and seasonal patterns in addition to the annual signal. We use a Bayesian PHM with spatial-temporal dependence, and add a prior that can facilitate the selection of areas with no association, taking into account other correlations among coefficients. 

Often the effect of covariates on the response is assumed to be constant across locations and time. However, this assumption is inappropriate in our case as TC association with SST and LHF features varies with strength, location and time within a season (\cite{Hodges2014} and \cite{Xie2014b}). Thus, we use spatial-temporally varying coefficients (\cite{Gelfand2003a}) that can relax this assumption. Moreover, our proposed method applies variable selection techniques in order to identify areas with significant association in the spatial-temporal domain.

Many variable selection methods have been proposed for Bayesian models; for example, stochastic search variable selection (SSVS) from \cite{George1993}. \cite{Smith2007} and \cite{Chipman1996} propose variations from SSVS, for single predictors and \cite{Scheel2013} propose to assume the coefficients of multiple predictors of interest as a product of a binary spatially dependent field and an independent Gaussian field, to perform spatial variable selection. \cite{Reich2010}, \cite{Lum2012} and \cite{Cai2013} describe SSVS specifically for spatially and spatial-temporally varying coefficients in Bayesian models. \cite{Reich2010} use the spike and slab prior from \cite{Ishwaran2005} and describe a computationally efficient way to estimate the model. \cite{Cai2013} propose a model that performs variable selection for spatial-temporally varying coefficients using a Dirichlet process prior and nonparametric techniques. Depending on the mixed distributions, the interpretation of these coefficients becomes very complicated and almost impossible. \cite{BoehmVock2015} propose to use copulas to retain interpretability of the coefficient. We propose a new version of a copula prior that can create a continuous smooth prior surface and perform multivariate spatial-temporal variable selection.

The novelty in our method is a dynamic variable selection procedure, which shares information between responses, locations, time and levels in the PHM context. This skill is essential in cases where we have a limited number of variables available, with a strong dependence structure, as in TC forecast models, but also in many other areas. For example, in epidemiology studies, where one is interested in detecting location and time where covariates have significant effects on a zero inflated response; or in forestry or ecology where models for sparse associations of covariates and rare events in space and time are also common. 

This paper is organized as follows. Section 2 describes the data used in the application. Section 3 introduces the statistical model and its estimation details. Section 4 describes a simulation study that investigates the performance of the variable selection technique, and Section 5 presents the results from real data application of the model. Finally, Section 6 presents our discussion and conclusions. 

\section{Data}

\begin{figure}
  \begin{center}
    \includegraphics[scale=0.53]{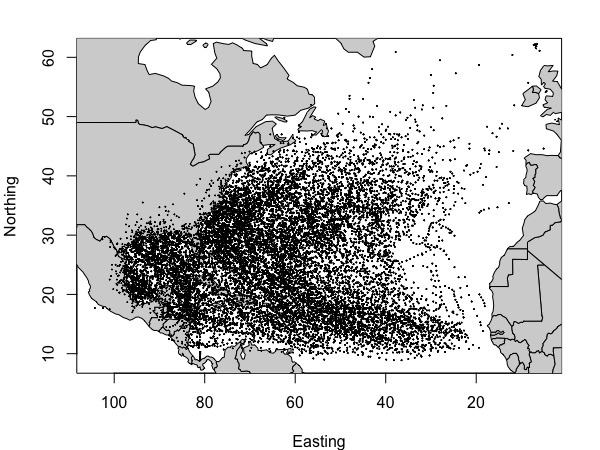}
  \end{center}
\caption{HURDAT 6 hour track data from 1950 to 2013}
\label{fig:mapall}
\end{figure}

Our main source of data is HURDAT (HURicane DATa), a publicly available database that contains detailed information for all recorded TCs in the Atlantic Basin since 1851. We construct a grid cell count using the 6 hour track data from 1950 to 2013 presented in Figure~\ref{fig:mapall}. The polygon grids are created using rectangles of 2.5 by 2.5 degrees over the Atlantic Basin, and using the coordinates from other variables used in the model as centroids. 

In meteorology, the 5 category Saffir-Simpson scale is used to describe hurricanes. In practice, forecasters report tropical cyclones in three strength groups according to said scale, which are:
\begin{itemize}
\item{Low}: Tropical storms. TCs with sustained winds of less than 33 $m/s$.
\item{Mid}: Hurricanes of category 1 and 2. TCs with sustained winds between 33 and 49 $m/s$.
\item{Strong}: Major hurricanes of category 3, 4 and 5. TCs with sustained winds of more than 50 $m/s$.
\end{itemize} We use this classification to count the number of unique storms per strength and year in each grid box. These counts are the response variable in our model.

\begin{figure}
\begin{center}
\includegraphics[width=0.75\textwidth]{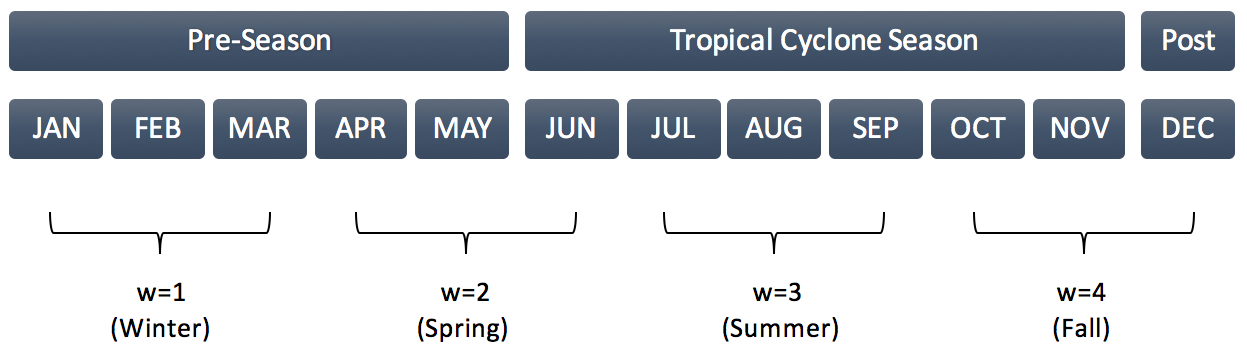}
\caption{Definition of trimesters compared to the Tropical Cyclone Season for TC data.}
\label{fig:seasons}
\end{center}
\end{figure}

\begin{figure}
  \begin{center}
    \includegraphics[width=1\textwidth]{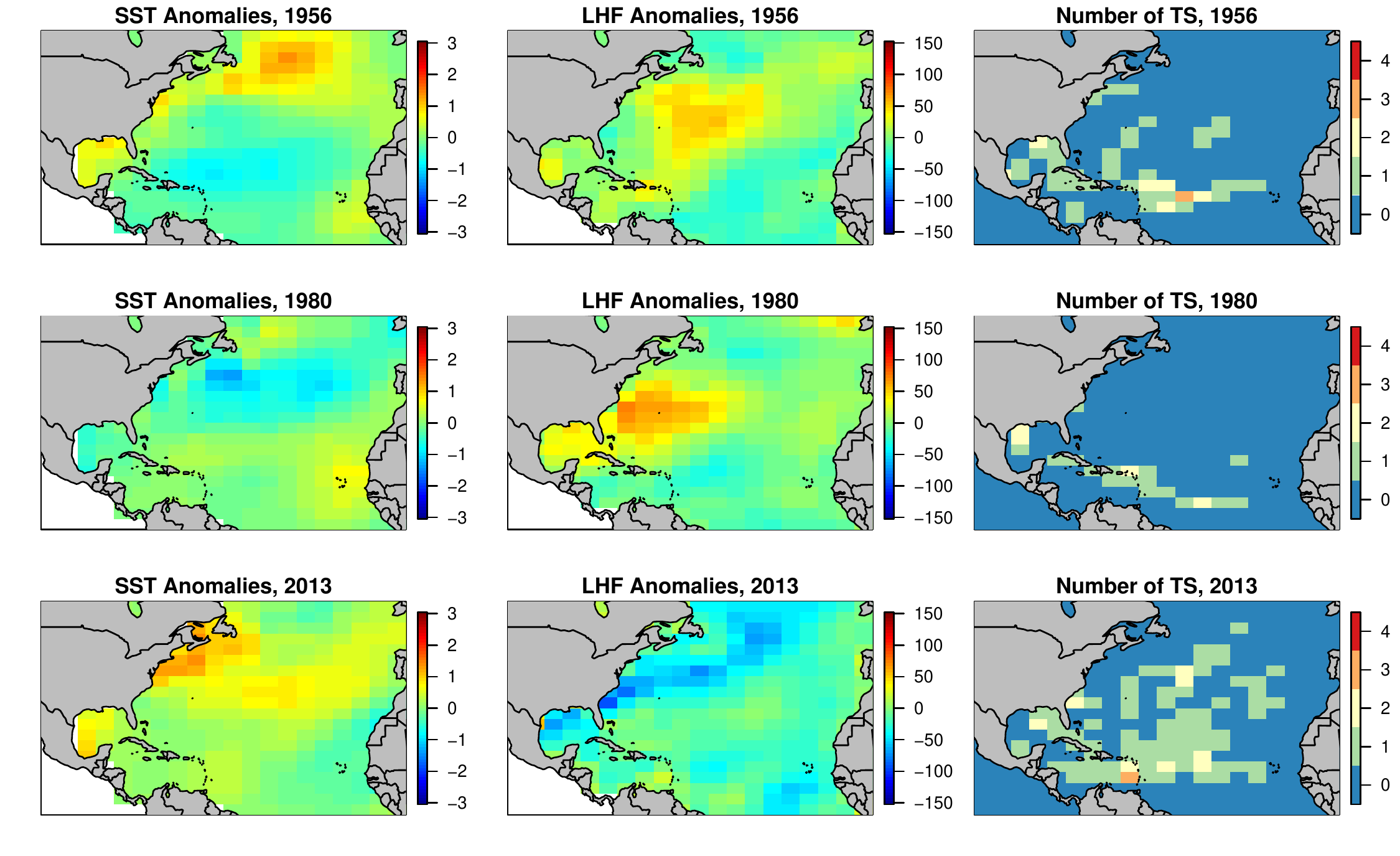}
  \end{center}
\caption{North Atlantic Basin: Anomalies in the North Atlantic Basin during the first trimester for SST (left column) and SST (middle column). Counts per grid box of Tropical Storms of Strength 1 (right column). Years 1955, 1980, and 2013 randomly selected to illustrate data.}
\label{fig:anomalies}
\end{figure}

SST and LHF are used as covariates in our model. For SST, we use data from the High-resolution Blended Analysis (\cite{Kalnay1996}), which takes measurements from NOAA's Advanced Very High Resolution Radiometer (AVHRR). SST is reported at a resolution of 2.5 by 2.5 degrees. LHF is defined as the flux of heat in watts per square meter ($W/m^2$) from the Earth's surface to the atmosphere, which is associated with evaporation of water at the surface and subsequent condensation of water vapor in the troposphere. We use data from the National Centers for Environmental Prediction (NCEP) Reanalysis (\cite{Kalnay1996}), with a 2.5 by 2.5 degree resolution. Both SST and LHF have daily observations available for the Atlantic Basin. We average the daily data over three-month periods (trimesters) that coincide with the definition of season from World Ocean Database (WOD) (Levitus (1983)), i.e. winter starting in January, spring in April, summer in July and fall in October. The relationship between trimesters $w$ and the tropical cyclone season is presented in Figure~\ref{fig:seasons}. We define $Z_{\ell}(s_i,t,w)$ as the average of covariate $\ell$, in trimester $w$, location $s_i$ and year $t$, and then anomalies $X_{\ell}(s_i,t,w)$ as departures from the temporal mean per trimester and location ($\overline{Z}_{\ell}(s_i,\cdot,w)$) in each variable. Anomalies for some selected years are presented in Figure~\ref{fig:anomalies}.

Our data consist of observations for $T = 64$ years, $M=4$ trimesters per year, $N = 400$ locations, $K = 3$ response variables (number of tropical cyclones of low, mid and strong strength), and $L = 2$ covariates: SST and LHF. Figure~\ref{fig:anomalies} presents maps of selected years as examples of the spatial locations in the Atlantic. 

\section{Statistical Methods}
In this section we describe our model for yearly counts of TCs per location and strength. The model simultaneously associates the different count distributions to covariates' features derived from anomalies.

\subsection{Pre-processing}
Since the association between SST and LHF and tropical cyclones is not necessarily limited to an association between measures in the same location, we seek multivariate spatial-temporal features to be used as covariates in an interpretative way, while still providing a practical fit to the data. We use Empirical Orthogonal Functions (EOFs) to describe said features. According to \cite{Lorenz1956}, EOFs are spatial components that display space-time modes of variability of a quantity over a region. Thus, each covariate anomaly, for each trimester $w$, ${\bf{X}}_{\ell}(w)=X_{\ell}(\cdot,\cdot,w)$ is approximated by the principal component (PC) scores, calculated using singular value decomposition (SVD) of ${\bf{X}}_{\ell}(w)$, for each $w$ and $\ell$. In this way, we can represent the anomalies as: 
\begin{equation}\label{eof}
X_{\ell} (s_i, t,w)=\sum\limits_{r=1}^{R} \xi_{\ell,r}(t,w)\phi_r (w,s_i) + \epsilon(s_i,t,w),
\end{equation}
where R is the number of scores used to explain more than $70\%$ of its variation. EOFs $\phi_r (w,s_i)$ are defined for each trimester $w$, $\xi_{\ell, r}(t,w)$ are the correspondent PC scores for each covariate $\ell$, and trimester $w$, and $\epsilon(s_i,t,w)$ is the variation orthogonal to the first R EOFs.

\subsection{Model}

Let $y_{k}(s_i,t)$ be the number of unique TCs in year $t$ of strength $k$ in location $s_i$. We model $y_{k}(s_i,t)$ using a hierarchical Poisson hurdle model (PHM), where the probability mass function is defined as:
\begin{equation} \label{eqPHM}
P( y_k(s_i,t) = m | p_k(s_i,t),\lambda_k(s_i,t) ) =  \left \{  \begin{tabular}{lr}
$1 - p_{k}(s_i,t)$& if $m = 0$ \\
$p_{k}(s_i,t) \frac{\lambda_{k}(s_i,t)^{m}}{(\exp^{\lambda_{k}(s_i,t)}-1) m!}$ & if $m > 0$\\
\end{tabular}   \right.
\end{equation} where $p_k(s_i,t)$ is the probability of non-zero counts and $\lambda_{k}(s_i,t)$ is the intensity parameter. Both $p_k(s_i,t)$ and $\lambda_{k}(s_i,t)$ are described as a logit and loglinear function of multivariate spatial-temporally varying coefficients of the covariates' scores, respectively. Thus we have:

\begin{equation} \label{eqLM}
\begin{split}
logit(p_{k}(s_i,t)) &=  \sum\limits_{\ell=1}^L \sum\limits_{w=1}^M \sum\limits_{r=1}^R \beta_{1,k,\ell,r}(s_i,w) \xi_{\ell,r}(t,w) \\
log(\lambda_{k}(s_i,t)) &= \sum\limits_{\ell=1}^L \sum\limits_{w=1}^M \sum\limits_{r=1}^R  \beta_{2,k,\ell,r}(s_i,w)\xi_{\ell,r}(t,w).
\end{split}
\end{equation}

Since our model coefficients vary in space by trimesters and by strength they are able to explain the association between the space-time modes of variability in the covariates (LHF and SST) and both the probability of non-zero counts and the intensity, for every strength. This association is expected to be close to zero in many spatial locations and during some trimesters, depending on the TC strength. To this end, we propose a method that can identify where, when, and for which strength this association is different from zero. Furthermore, we present a prior for $\beta$ that selects such cases and uses the correlation between responses, locations, time and levels in the PHM context to improve this estimation.

\subsection{Multivariate Spatial-Temporal Prior} \label{sec:prior}

Let $A$ be a combination of covariate $\ell$, PC score $r$, response $k$ and level of PHM $j$. Each coefficient $\beta_{A}(s_i,w)$ is modeled using a Gaussian copula and a latent variable  $\theta_{A}(s_i,w)$. We define $\theta_{A}(s_i,w) \sim \text{MVN}(0,\Sigma)$ and use $\Sigma$ with a separable covariance structure 
with the following form:
\begin{equation}\label{eqgamma}
\Sigma_{\theta}= \Gamma_s \otimes \Gamma_w \otimes \Gamma_k \otimes \Gamma_j.
\end{equation}
The term separable means that we can factor the covariance structure into purely spatial ($s$), temporal ($w$), multivariate ($k$), and by PHM level ($j$) covariance structures, which allows for computationally efficient estimation and inference. Each covariance $\Gamma$ can be written as a correlation function $C(h)$ where $h$ is the Euclidean distance when the distance is defined in space ($s$) or in time ($w$), or the numerical absolute difference when the function is in terms of the ordered response strengths ($k= 1,2,3$) or PHM levels ($j=1,2$).

Then, we define $ \beta_{A} (s_i,w) = F^{-1}_A \{\Phi[\theta_{A}(s_i,w)]\} $ where $\Phi$ is the cdf of a standard normal distribution and $F_A$ is the marginal cdf of each $\beta_{A}(s_i,w)$, distributed as a multivariate spike and slab mixture of normal densities: 
\begin{equation} \label{eqMP}
\beta_{A}  \sim \pi_{A} N(\alpha_{A}, \sigma^2_{A}) + (1-\pi_{A})N(0, \sigma^2_{A}/C), 
\end{equation}
where $\pi_A$ represents the expected proportion of coefficients that are centered at $\alpha_A$. In this way, coefficients are smoothed around $\alpha_A$ or 0. $\sigma^2_A$ controls variability around the means of the mixed distribution, and C represents the ratio of variance for coefficients centered in zero to centered in $\alpha_A$.

A copula can flexibly model the marginal distributions while introducing dependence via multivariate normal distributions, creating a continuous smooth prior surface. The intuition in our case is that if a coefficient is close to zero in a specific location and time, nearby locations (in space and time) are unlikely to have large coefficients. Also, coefficients for adjacent categories from an ordered response and the two levels of a PHM are expected to be correlated, since the probability of having zero counts in a specific location and time is likely to be associated with its intensity.

For $\pi_A$, we use a flat prior between 0 and 1. Other prior choices can give more weight to options like $\pi_A=0$: all locations and trimesters have a coefficient centered in zero for $A$, or the opposite $\pi_A=1$. For $\alpha_A$, the prior is a normal distribution with standard deviation 1, which allows the coefficients to be centered in values ranged from $[-4,4]$. Previous work on mixed models similar to Equation~\ref{eqMP} has defined that model fit can be sensitive to large values of $\sigma_A$ because it would be difficult to differentiate between coefficients centered at 0 from other values. We use a Gamma(0.1,0.1) prior to encourage small values for $\sigma_A$.

We model the temporal and spatial covariance components using AR1 and exponential structures, respectively, while using a vague Wishart prior to model the multivariate covariance and the covariance between PHM levels. Hyperpriors for the spatial range parameter $r$, and the temporal correlation and variance $\rho_t$ and $\sigma_t$ are recommended to be informative, in the sense of restricting the range of possible parameter values to values which make sense in terms of the application.

To estimate the parameters in the model, we use a Markov chain Monte Carlo (MCMC) algorithm. Since we are using a mixed distribution as a prior for $\beta_{A}(s_i,w)$, we do not have conjugacy, and we need to use rejection sampling. The scheme requires factoring $24N \times 24N$ matrices that we simplify by using the Woodbury matrix identity on the Kronecker structure of $\Sigma_{\theta}$. In the simulation study in Section 4 we generate 15000 iterations, discard the first 5000 as burn-in, and thin the chains by keeping every fifth iteration.

\section{Simulation Study}

We study the performance of our model compared to two other competing models for variable selection. Our main focus is to describe the dynamic variable selection performance; thus we compare performance for varying degrees of spatial-temporal correlation in the coefficients $\beta_A(s_i,w)$. For that, we use the following statistics: median absolute deviation (MAD), ability to correctly identify null coefficients in space and time and MSE for each coefficient estimated. 

We generate $5$ $\beta_{A} (s_i,w)$ under three different settings. Setting 1 is composed of smooth surfaces created using linear combinations of latitude, longitude, and $\alpha_A= \{2,1.5,1,0.5,0\}$, respectively. Figure 1 in \ref{suppB} shows an example realization. In this case $\beta_{A} (s_i,w)$ does not follow the same distribution as in any of our models. Settings 2 and 3 are generated using the copula transformation described in Section~\ref{sec:prior} and assuming~\ref{eqMP} true, with $\alpha_A= \{2,1.5,1,0.5,0\}$, $\pi=\{1,0.8,0.5,0.2,0\}$, $\sigma^2_{A} = 1$ and $C= \infty$. They differ in the degree of spatial-temporal correlation. For each $A$ and each setting, we generate $M=4$ time points and $N=100$ locations. 

Samples of $\xi_{A} (w, t) \sim \text{N}(0, 2^{-1/2})$ for $w=1,2,..,4$ and $t = 1,2,..,30$, with $A =1,2,..,5$ are generated to simulate scores. We apply Gram-Schmidt orthonormalization to make them independent among $A$. Then, we use $\beta_{A} (s_i,w)$ from settings 1-3, to have $\boldsymbol{\beta}^T \boldsymbol{\xi}$, and use it to generate random samples of $Y(s_i,t)$ from a Poisson distribution: $\text{Pois}(\exp{(\boldsymbol{\beta}^T \boldsymbol{\xi})})$.

Following the procedures outlined above, we generate B = 50 data sets using the following settings for $\beta_{A} (s_i,w)$:

\begin{itemize}
\item {\bf{Setting 1: }}Weak Temporal Correlation and Strong Spatial Correlation. Smooth surfaces are generated using linear combinations of latitude and longitude.
\item {\bf{Setting 2: }}Weak Spatial-Temporal Correlation. The true coefficients for each location and trimester are exponentially spatially correlated with range = 2 km and temporally correlated with $\rho=0.1$ for all A.
\item {\bf{Setting 3: }}Strong Spatial-Temporal Correlation. The true coefficients for each location and trimester are exponentially spatially correlated with range = 100 km and temporally correlated with $\rho=0.9$ for all A.
\end{itemize}

We test the performance of three models under three different data settings. Since we concentrate on showing the advantages of sharing information between locations and times when performing variable selection, we simplify the response model to be $y(s_i,t) \sim \text{Pois}(\lambda(s_i,t)),$ and use the second Equation in~\ref{eqLM} to model the intensity parameter $\lambda(s_i,t)$ as a function of the scores $\xi_{\ell,r}(w,t)$. Here, $A$ represents scores for each covariate (component) that is used in our model. The three models we consider are our model ({\bf{Model 3}}) and two other competing models for variable selection. They differ in the latent variable distribution used in the transformation for $\theta_{A}(s_i,w) \sim \text{MVN}(0,\Sigma_{\theta})$: 

\begin{itemize}
\item {\bf{Model 1: }}$\beta_A(s_i,w)$ are independent across sites and times and follow the distribution in~\ref{eqMP}. It does not share information, since $\Sigma_{\theta} = \sigma^2_{t} \times I$ is just a diagonal matrix.
\item {\bf{Model 2: }}Spatial dependence in $\beta_A(s_i,w)$ is induced with a Gaussian Copula. It shares information across sites, using $\Sigma_{\theta} = S \otimes  \sigma_t^2 \times I$ where $S$ represents an exponential spatial correlation matrix with range r.
\item {\bf{Model 3: }}Spatial-temporal dependence in $\beta_A(s_i,w)$ is induced using a Gaussian Copula. It shares information across sites and times, using $\Sigma_{\theta} = S \otimes T$, where S is defined as in Model 2 and T has an autoregressive 1 structure (AR1) with parameters $\rho_{t}$ and $\sigma^2_{t}$ for time. 
\end{itemize}

In all cases, we use the same priors : $\alpha_A \sim \text{N}(0,2)$, $\pi_{A} \sim \text{Beta}(1,1)$, and $\sigma^2_A \sim \text{InvGamma}(0.1,0.1)$. We include Model 1 as a baseline to show the overall improvements in the estimation of $\beta_{A} (s_i,w)$ when considering the spatial and the spatial-temporal structure that models 2 and 3 provide. 

We compute posterior medians and credible intervals for each $\beta_{A}(s_i,w)$ and posterior means for $\pi_A$ and $\alpha_A$. Median absolute deviation (MAD) is averaged over space, trimesters and iterations for each $\beta_{A}$. Percentage of correctly estimated zeroes are calculated only for $\beta_{A}$ and averaged over space, trimesters and iterations as well. MSE is also calculated and averaged over iterations for $\alpha_A$ and $\pi_A$.

MAD is computed in the following way, for each setting and model:
\[
\text{MAD}_{A} = \text{Median}_{\mathit{BNM}} |\hat{\beta}_{A}(s_i,w) - \beta_{A}(s_i,w)|
\] where $B = 50$ data sets, $N = 100$ locations and $M = 4$ time points. 

\begin{table}
\footnotesize
\centering
\caption{Mean Acceptance Rate for all parameters.}
\begin{tabular}{rrrrrrrrrr}
\hline
\hline
& \multicolumn{3}{c}{Weak T Correlation}  & \multicolumn{3}{c}{Weak ST Correlation}  & \multicolumn{3}{c}{Strong ST Correlation}  \\
 & M1 & M2 & {\textcolor{red}{M3}}  & M1 & M2 & {\textcolor{red}{M3}} & M1 & M2 & {\textcolor{red}{M3}} \\
  \hline
\hline
$\beta$ & 0.75 & 0.74 & 0.44 & 0.81 & 0.79 & 0.76 & 0.74 & 0.79 & 0.43 \\ 
$\alpha$ & 0.19 & 0.14 & 0.35 & 0.71 & 0.71 & 0.85 & 0.71 & 0.71 & 0.85 \\ 
$\pi$ & 0.31 & 0.35 & 0.38 & 0.36 & 0.35 & 0.32 & 0.35 & 0.35 & 0.43 \\ 
 \hline
\end{tabular}
\label{tab:ACC}
\end{table}

Table \ref{tab:MAD1} presents MAD for each $\beta_{A}(s_i,w)$. Values were compared using credible intervals. We can see how the median absolute deviations for Model 3 are statistically smaller in most of the settings generated with strong correlation, as expected. In some cases, this improvement is reflected in a $40\%$ decrease of MAD. Nevertheless, this difference is not reflected in cases where we have weak correlation and $\alpha$ is generated as far from zero ($\beta_{1}$ and $\beta_{2}$). When comparing results for data generated with weak temporal correlation, MAD for our model is smaller than models 2 and 3 in all cases. MAD is statistically smaller using Model 3 in cases where the real $\alpha$ is equal or close to zero ($\beta_{3}$,  $\beta_{4}$, and $\beta_{5}$).

\begin{table}
\centering
\caption{Median MAD per $\beta_A$ over locations and trimesters. All standard errors are less than 0.006. Proportion of correctly estimated $\beta_{A}=0$  All standard errors are less than 0.005. NA in cases where we did not simulate zeroes.}
\begin{adjustbox}{width=1\textwidth}
\footnotesize
\begin{tabular}{rrrrrrrrrrr}
 \hline
\hline
&& \multicolumn{3}{c}{Weak T Correlation}  & \multicolumn{3}{c}{Weak ST Correlation}  & \multicolumn{3}{c}{Strong ST Correlation}  \\
 && M1 & M2 & {\textcolor{red}{M3}}  & M1 & M2 & {\textcolor{red}{M3}} & M1 & M2 & {\textcolor{red}{M3}} \\
\hline
\hline 
MAD&$\beta_{1}$& 0.3073  & 0.4018 & 0.3699 & 0.6235  & 0.5996  & 0.7316 & 0.7594  & 0.6712 & 0.6882 \\ 
&$\beta_{2}$ &  0.3229  & 0.4006  & 0.3686 & 0.6368  & 0.5843  & 0.7327 & 0.6731  & 0.6787  & 0.6112 \\ 
&$\beta_{3}$ & 0.3263  & 0.3440  & 0.2824 & 0.6940  & 0.7460  & 0.4722 & 0.9846  & 1.0040  & 0.5306 \\  
&$\beta_{4}$ & 0.3349  & 0.3380  & 0.1999 & 0.5110  & 0.5174  & 0.3615 & 0.9221  & 0.8880  & 0.3044 \\ 
&$\beta_{5}$ & 0.3197  & 0.3268  & 0.2042 & 0.6480  & 0.6458  & 0.0154 & 0.7164  & 0.7049  & 0.2473 \\ 
\hline
\hline
Proportion&$\beta_1$ & 0.8121  & 0.7437  & 0.8862 & NA &  NA& NA&  NA & NA &  NA \\ 
 of Correctly&$\beta_2$ & 0.8457  & 0.7594  & 0.8670 & 0.5258  & 0.4419  & 0.7860 &   0.7792  & 0.8430 & 0.8404 \\ 
Estimated &$\beta_3$ & 0.7434  & 0.8369  & 0.9409 & 0.8925  & 0.8754  & 0.9437  & 0.6028  & 0.7602  & 0.9692 \\ 
$\beta_{A}=0$ &$\beta_4$ & 0.7184  & 0.7346  & 0.8676 & 0.9586  & 0.9746  & 0.9894 & 0.8568  & 0.9476  & 0.9827  \\ 
&$\beta_5$ & 0.8941  & 0.8541  & 0.9285 & 0.5510  & 0.5509  & 0.9980 & 0.5563  & 0.5517  & 0.8882 \\ 
\end{tabular}
\end{adjustbox}
\label{tab:MAD1}
\end{table}

As a way to evaluate the performance of our dynamic variable selection method, we calculate the proportion of correctly estimated $\beta_{A}=0$.  Our model detects a statistically higher proportion of real values equal to zero in all cases when the data is generated with weak and strong correlation. On the other hand, our model performs similarly to Model 2 in cases where we have weak temporal correlation or when $\alpha$ is far away from zero. Simulation results show an average increase of $13\%$, $38\%$, and $30\%$ in the percentage of correctly detected zeroes, for each of the settings. The best case scenario of performance for our dynamic variable selection procedure is when data presents strong spatial-temporal correlation, and it is close to zero.

Intuitively, MAD result reflects the fact that our model (M3) benefits from using temporal correlation to get a better fit when there is strong spatial-temporal correlation present, compared to the fit to weak spatial-temporal or temporal dependent data. The same benefit is reflected in the percentage of correctly detected zeroes, where sharing information in space and time gives the model more tools to correctly detect zeroes. Therefore, when there is strong spatial-temporal correlation present in the data, the simulation shows the need of a model that accounts for both space and time dependence.
 
Overall, our model shows an average increase of $13\%$, $38\%$, and $30\%$ in the percentage of correctly detected zeroes for weak temporal, weakly and strongly correlated data, respectively. Also, on average our model has a MAD per $\beta_A$ $38\%$ lower than the competing models for strongly correlated data, $24\%$ lower for weakly correlated data, and $16\%$ lower with weak temporal correlation. 

\section{Analysis of Tropical Cyclones in the Atlantic Basin}  

\subsection{Model Comparison}

We fit the full model (ST) from Section 3, and two other models (Spatial and Independent) to the tropical cyclone data described in Section 2. The difference between the full model (ST), spatial (SP) and independent (IN) model is the specification of $\Sigma_{\theta}$ in \ref{eqgamma}. In the SP model we use $\Sigma_{\theta}= \Gamma_s \otimes I_w \otimes I_k \otimes I_j$, and in the IN model we assume independence over all levels ($\Sigma_{\theta}= I_s \otimes I_w \otimes I_k \otimes I_j$). We perform the comparison in order to collect evidence of the need of the full model (or the lack thereof).

For each trimester $w$, we calculate anomalies $X_{\ell}(s_i, t,w)$ for both SST and LHF and then perform a SVD as described in Section 3. We obtain R = 4 EOFs with their respective scores $\xi_{\ell, r}(t,w)$ for each covariate $\ell$ and trimester $w$, to have 16 time series in total. The maps of the four resulting EOFs for each covariate and trimester are presented in the~\ref{suppB}. The obtained EOFs for each of the seasons and covariates together describe at least $70\%$ of the variability. The estimated scores range from $-6$ to $6$ in all cases and are centered close to zero. In the results subsection we describe these spatial patterns as features and relate them to the response with the statistically significant factors. 

We compare models using DIC and MSE, to evaluate goodness of fit of the model, but also their ability to estimate correctly the number of storms per location and per year. The percentage of locations with no TCs in a given year in the Atlantic Basin ranges from $78\%$ to $98\%$ for low strength, from $83\%$ to $99\%$ for mid strength and from $89\%$ to $100\%$ for strong strength and the range of counts per location is from 0 to 4 storms. Table~\ref{tab:bicmse} presents all the goodness of fit measures for each model. 

\begin{table}
\begin{center}
\caption{Goodness of fit criterion DIC, and MSE overall, per location and per year for the full model (ST), spatial (SP) and independent (IN). DIC bar values (standard error in parenthesis) for models using different grid box sizes, compared to a model using climate indices.}\label{tab:bicmse}
\begin{tabular}{c|c|ccc}
&& \multicolumn{3}{c}{MSE}  \\
	 &$\overline{DIC}$ & TS & HU & MH \\
\hline
Independent (IN) &183.1 (1.2)  &3.16 (0.3) & 2.51 (0.2) &0.08 (0.001) \\
Spatial (SP) &142.9 (1.1)&0.20 (0.01) & 0.18 (0.02) &0.04 (0.001) \\
Spatial Temporal (ST) & 136.2 (4.2)&0.13 (0.01) &0.12 (0.001) &0.03 (0.001) \\
\hline
200 grid boxes & 136.9 (4.5)&&& \\
400 grid boxes & 136.2 (4.2)&&&\\
800 grid boxes & 136.5 (4.3)&&& \\
Climate indices & 173.6 (5.8)&&& \\
\hline
\end{tabular}
\end{center}
\end{table}

The full model (ST) is selected as the best model by DIC and MSE values for every response. We compare the observed with the expected values of the posterior distribution and obtain that the full model performance in terms of MSE per response is superior to the IN and SP models performance for all responses. Figures~\ref{fig:MSEyear} and~\ref{fig:MSEloc} in the Appendix show the MSE per location and year, respectively for the full model. Our model has a fairly superior performance per year compared to climatology (MSE per response is $0.14$ for TC, $0.16$ for HU and $0.03$ for MH), with some exceptions like 1966, 1997 and 2005 for HU. In addition, the full model presents MSE lower than 1 unit in most of the locations, with a unique exception in front of the African Coast, very close to the mid TC main development region. However, prediction is not the only focus of our model. We also use the model to identify potentially scientifically relevant associations between covariates and responses.

\subsection{Sensitivity Study}

We consider several choices of hyperpriors when fitting the full model. Also, we estimate the scores using three different rectangle sizes as grid boxes in space for SST and LHF measures. We evaluate the performance of the different modeling options using the Deviance Information Criterion (DIC). 

We specify different hyperpriors for the spatial range parameter $r$, and the temporal correlation and variance $\rho_t$ and $\sigma_t$. Results were robust in terms of DIC when changing priors for $\rho_t$ and $\sigma_t$ from informative to non informative. For this analysis we use a uniform prior for $\rho_t$ and a Gamma(0.1,0.1) for $\sigma_t$. We observe that DIC is sensitive to a change in $r$; goodness of fit improves significantly when using a uniform distribution from $0$ to $300$ km, instead of $30$ km or $3000$ km as upper limits. This improvement could be explained by the typical ratio of a tropical cyclone eye, which approximately ranges from $3$ to $300$ km (\cite{NHC2016}). 

In addition to these hyperprior choices, we conduct a specific sensitivity analysis to compare the goodness of fit when using different rectangle grid box sizes to construct the SST and LHF scores, and at the same time compare it with a model that uses climate indices instead. The use of climate indices as base comparison is particularly interesting in this case, since those are the covariates that are typically used in TC forecast models (\cite{Xie2014}). DIC was nearly identical among the cases of using 200, 400 and 800 grid boxes to construct the SST and LHF scores, but it was higher in the model that uses climate indices instead of scores (Table~\ref{tab:bicmse}). We set the rectangle size to be the same as the response (2.5 by 2.5 degrees) and use 400 grid boxes to calculate the SST and LHF scores. The first two scores per trimester are presented in~\ref{suppB}.

In order to improve MCMC convergence, we tune $C$ to give acceptance rates of around $30\%$ or higher for most parameters. Convergence is monitored using trace plots of deviance and several representative parameters. For all the sensitivity analyses and the final analysis, we generate 20,000 samples and discard the first 5,000 as burn-in and thin the chains by keeping every fifth iteration. 

\subsection{Results}

Table~\ref{tab:RES2} presents the posterior estimates for each response for the overall mean $\alpha_A$, the proportion $\pi_A$ of coefficients centered in $\alpha_A\neq 0$ and  the variability $\sigma_A$ of coefficients centered in $\alpha_A$. The proportion of locations where estimated coefficients are significantly different from zero for SST and LHF features are presented in Table~\ref{tab:RES1}; significant difference is determined by whether or not credible intervals include zero. 

\begin{table}
\centering
  \small
\caption{Posterior estimates for each response for the overall mean $\alpha_A$, the proportion $\pi_A$ of coefficients centered in $\alpha_A\neq 0$ and  the variability $\sigma_A$ of coefficients centered in $\alpha_A$, for responses $Y_1 =$ Tropical Cyclones, $Y_2 =$ Hurricanes and $Y_3 =$ Major Hurricanes. Values with (*) have credible intervals that do not include zero.}
\begin{adjustbox}{width=1\textwidth}
\begin{tabular}{r|rrrr|rrrr|rrrr}
  \hline
 & \multicolumn{4}{c}{Tropical Cyclones} &\multicolumn{4}{c}{Hurricanes} &
\multicolumn{4}{c}{Major Hurricanes} \\
 & SST, $p$ & LHF, $p$ & SST, $\lambda$ & LHF, $\lambda$& SST, $p$ & LHF, $p$ & SST, $\lambda$ & LHF, $\lambda$& SST, $p$ & LHF, $p$ & SST, $\lambda$ & LHF, $\lambda$ \\ 
\hline
$\alpha_A, r=1$ & -0.44 (*)& 0.56 & 0.13 & 0.18 (*) & 0.01 & -0.02 & 0.11 & 0.29 &  -0.02 & 0.50 (*)& 0.09 & 0.46 (*)\\ 
$\alpha_A, r=2$ & -0.49 (*)& -0.48 (*)& -0.71 (*)& -0.21 & -0.40 (*)& -0.36 & 0.09 & -0.31 &   0.17 & -0.33 (*)& 0.04 & -0.39 (*)\\ 
$\alpha_A, r=3$ & -0.76 (*)& 2.99 (*)& -0.59 (*) & 0.24 (*)& -0.31 & 0.08 & -0.16 & -0.50 &  -0.38 & -0.48 (*)& -0.18 & -0.76 \\ 
$\alpha_A, r=4$ & -0.64 (*)& -0.34 (*)& -0.03 & -0.41 (*)& 0.26 & 0.01 & 0.13 & 0.15 &   -0.07 & -0.00 & -0.07 & 0.09 \\
\hline
$\pi_A, r=1$ & 0.19 (*)& 0.16 (*)& 0.01 & 0.08  (*)& 0.03 & 0.52 (*)& 0.12 (*)& 0.15 (*) & 0.04 & 0.18 (*) & 0.09 (*)& 0.02 \\ 
$\pi_A, r=2$ & 0.22 (*)& 0.28 (*)& 0.33 (*)& 0.80 (*)  & 0.22 (*)& 0.06 & 0.19 (*)& 0.17  (*)& 0.05 & 0.28 (*)& 0.22 (*)& 0.05 \\ 
$\pi_A, r=3$ & 0.21 (*)& 0.14 (*)& 0.33 (*)& 0.67 (*)  & 0.04 & 0.04 & 0.13 (*)& 0.19 (*)& 0.25 (*)& 0.23 (*)& 0.05 (*)& 0.02 \\ 
$\pi_A, r=4$ & 0.24 (*)& 0.21 (*)& 0.01 & 0.18 (*) & 0.17 & 0.55 (*)& 0.12 (*)& 0.17  (*)& 0.07 & 0.05 & 0.21 (*)& 0.16 (*)\\
\hline
$\sigma_A, r=1$ & 0.10 (*)& 0.09 (*)& 0.05 (*)& 0.05 (*) & 0.11 (*)& 0.51 (*)& 0.48 (*)& 0.64 (*)& 0.17 (*)& 0.21 (*)& 0.49 (*)& 0.09 (*)\\ 
$\sigma_A, r=2$  & 0.13 (*)& 0.20 (*)& 0.08 (*)& 0.83 (*) & 0.17 (*)& 0.11 (*)& 0.61 (*)& 0.49  (*)& 0.16 (*)& 0.30 (*)& 0.70 (*)& 0.38 (*)\\ 
$\sigma_A, r=3$ & 0.16 (*)& 0.20 (*)& 0.08 (*)& 1.34 (*) & 0.11 (*)& 0.11 (*)& 0.51 (*)& 0.87 (*)&  0.25 (*)& 0.28 (*)& 0.43 (*)& 0.21 (*)\\ 
$\sigma_A, r=4$  & 0.20 (*)& 0.08 (*)& 0.05 (*)& 0.05 (*) & 0.12 (*)& 0.59 (*)& 0.50 (*)& 0.52 (*)& 0.17 (*)& 0.17 (*)& 0.55 (*)& 0.57 (*)\\ 
 \hline
   \end{tabular}
   \end{adjustbox}
\label{tab:RES2}
\end{table}

\begin{figure}
\begin{center}
\includegraphics[width=1\textwidth]{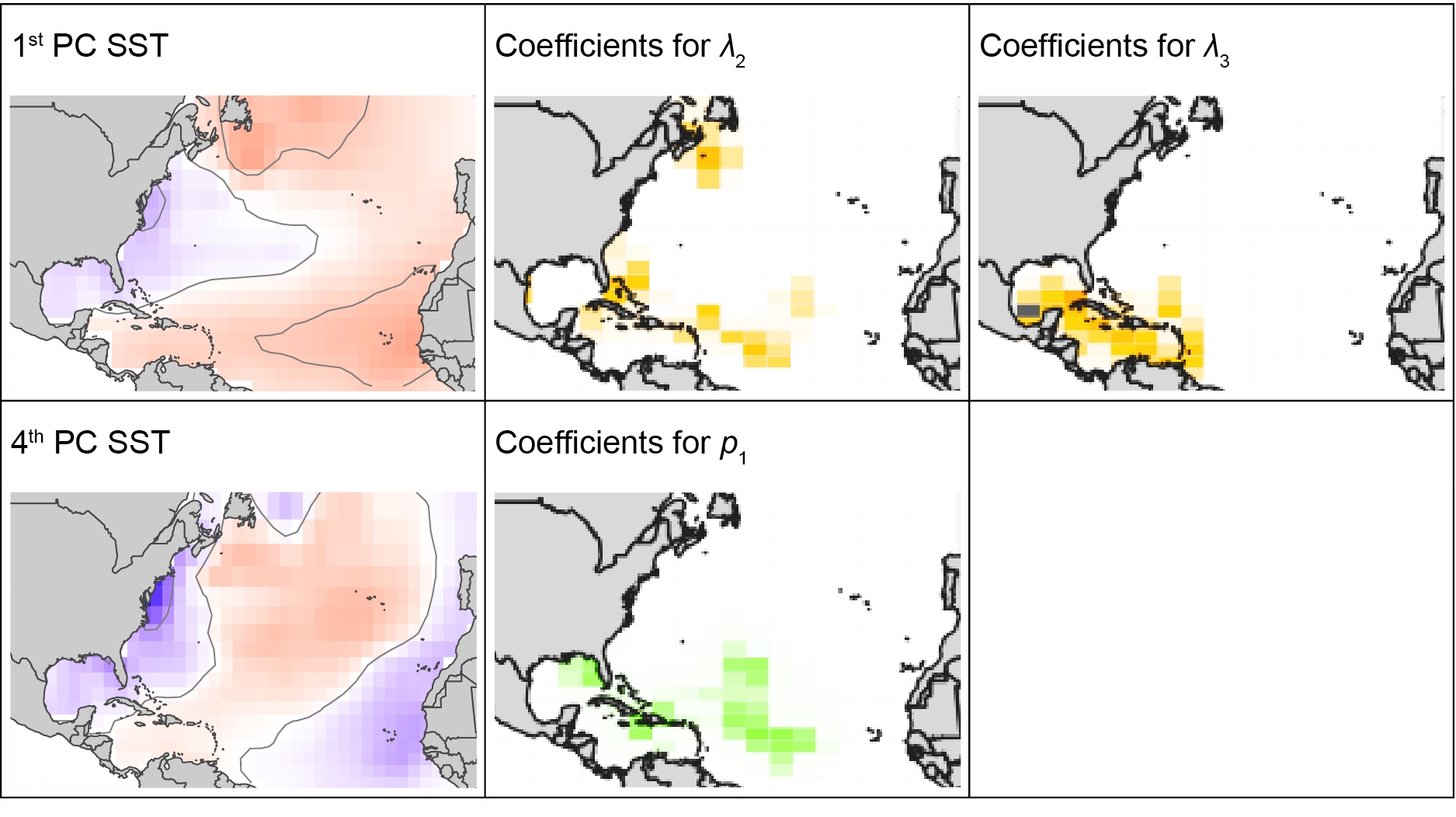}
\caption{Winter: SST Features (first left column) and their respective statistically significant associated factors. Warm colors (red, orange) are positive, and cold colors (purple, green) are negative. EOFs range from -0.2 to 0.2 and coefficients from -1 to 1.}
\label{fig:winter}
\end{center}
\end{figure}

We define a significant factor as an EOF for which more than $10\%$ of the corresponding spatially varying coefficients for a given trimester $w$ are statistically different from zero (posterior credible interval do not include zero).
We describe here those factors that have a sum of magnitudes greater than 20 for intensity and greater than 10 for probability of occurrence. SST features are the only significant factors of coefficients during winter ($w=1$). These coefficients are the most important in terms of determining predictors to use in the seasonal forecast, because they can be measured before the TC season starts (Figure~\ref{fig:seasons}). For example, the model indicates that warmer anomalies in SST over the mid TC main development region (in front of the African Coast) and in the Arctic during the winter are positively correlated with more major hurricanes in the Caribbean Sea and with more hurricanes around Florida, Cuba and the Mid Atlantic. Furthermore, the model shows that negative anomalies in SST over the Western Atlantic present a negative correlation with the occurrences of tropical storms over the coast of Florida and Cuba (see Figure~\ref{fig:winter}).

\begin{figure}[t]
\begin{center}
\includegraphics[width=1\textwidth]{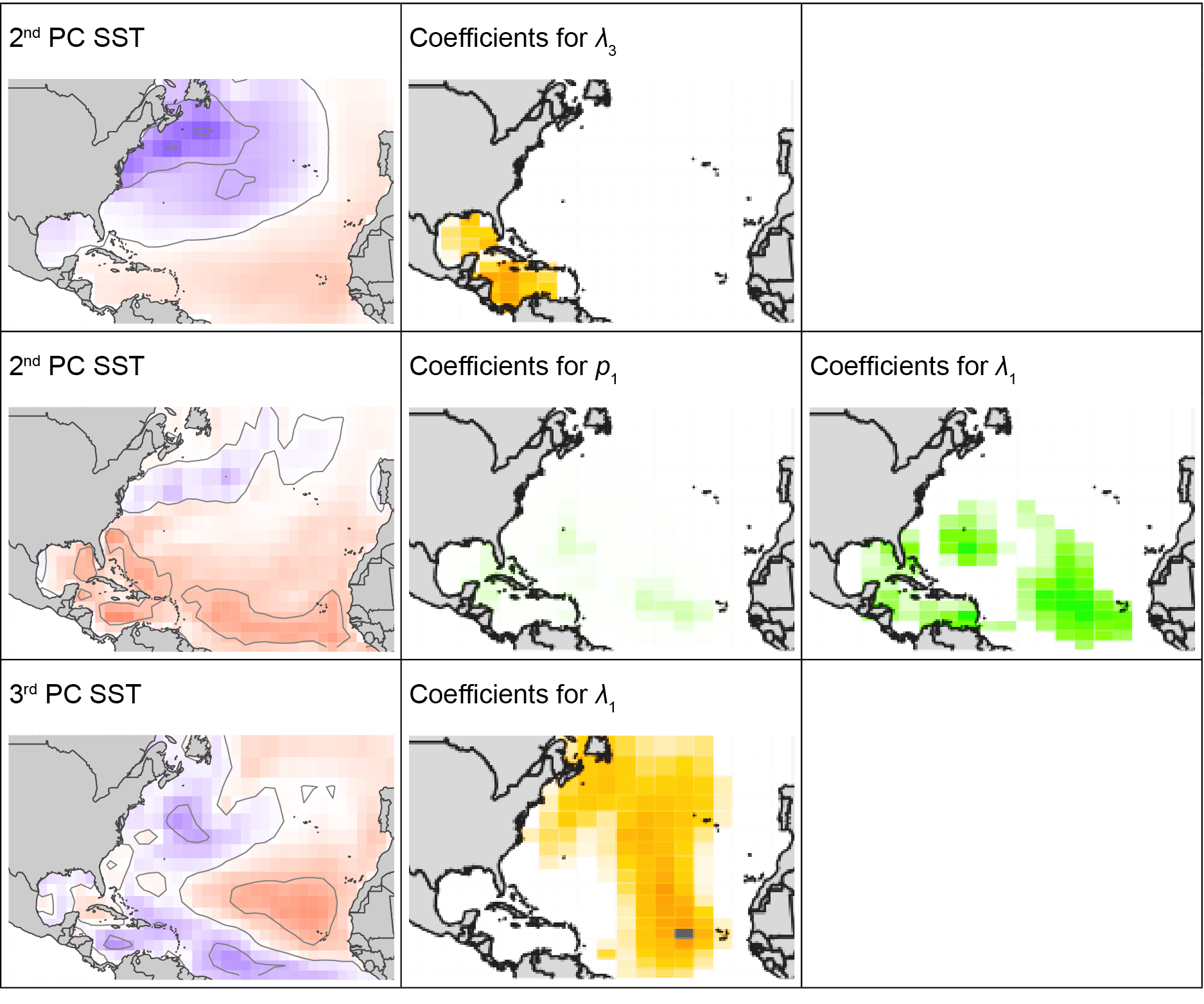}
\caption{Spring: SST or LHF Features (first left column) and their respective statistically significant associated fractors. Warm colors (red, orange) are positive, and cold colors (purple, green) are negative. EOFs range from -0.2 to 0.2 and coefficients from -1 to 1.}
\label{fig:spring}
\end{center}
\end{figure}

TC seasonal variation is well described using SST and LHF features during spring and summer ($w = 2,3$). While these covariates cannot be used for seasonal predictions, identifying their relationship with TC occurrence may provide guidance for future scientific studies. During spring, we have a significant factor that describes a positive correlation between positive anomalies in SST over the Arctic and the number of major hurricanes in the Gulf of Mexico and the Caribbean Sea. Also, Figure~\ref{fig:spring} shows three significant factors that relate positive anomalies of LHF over the tropical Atlantic, Caribbean and Gulf of Mexico as positively associated with both the occurrence of tropical storms and their number of occurrences over the Atlantic.

During summer (Figure~\ref{fig:summer} in the Appendix), we find an opposite pattern for SST anomalies in the Arctic, since they are negatively correlated with the number of hurricanes and major hurricanes over the Gulf of Mexico, around Cuba and over the Mid Atlantic. Furthermore, positive anomalies of LHF in front of the US East Coast are positively correlated with the occurrence of hurricanes in the same area. Finally, positive LHF anomalies over the Gulf of Mexico and the tropical Atlantic are negatively correlated with the occurrence of major hurricanes around the lower Antilles and with the number of tropical storms over the Gulf of Mexico, the Caribbean and the Mid Atlantic. Lastly, during the fall (Figure~\ref{fig:fall} in the Appendix) all the statistically significant factors are related to LHF. 

\section{Discussion} 

In this article, we develop a statistical model that allows for dynamic variable selection. The multivariate spatial temporal smooth surface we use in the prior reflects the fact that categories, trimesters, locations, and levels of PHM can affect the association of SST and LHF with the response. 

A simulation study shows that our model outperforms the independent and spatial models providing an average increase of $27\%$ in the percentage of correctly detected zeroes, in all treatments. Also, on average our model has a MAD per $\beta_A$ $26\%$ lower than the competing models. In some cases, this improvement is reflected in a $40\%$ decrease of MAD. In the data application, we find that using 200, 400 or 800 grid boxes to construct the SST and LHF scores has no effect on model goodness of fit (DIC), but each is better than a traditional climate index model (Table~\ref{tab:bicmse}). Also, our model has a significantly lower MSE values than the competing models. 

This study provides a better understanding of how observed TC variations in space and time are related to variations in features of SST and LHF. No previous study has attempted to relate SST and LHF features to the variation of storms, using spatial-temporal models and variable selection techniques. Our results show that warmer SST anomalies during winter and spring over the mid TC main development region are associated with more major hurricanes in the Caribbean Sea and the Gulf of Mexico. The results of this study can be used to improve seasonal forecast models by using the proposed measures as seasonal predictors.


\appendix

\section{Descriptive Statistics for Coefficients}\label{appD}
\begin{table}[H]
\centering
\caption{Proportion of locations with estimated coefficients statistically different from zero (credible intervals do not include zero). From those coefficients different from zero (+) indicates that more than $60\%$ of the locations have a positive coefficient for that specific trimester, response and score. In a similar way, (-) indicates that more than $60\%$ of the locations have a negative coefficient for that specific trimester, response and score. Bold values are significant factors. Red values are those displayed in Figures~\ref{fig:winter},~\ref{fig:spring},~\ref{fig:summer} and ~\ref{fig:fall}.} 

\begin{adjustbox}{width=1\textwidth}
\begin{tabular}{c|c|cccc|ccccc}
\hline
Resp, Level & Score & \multicolumn{4}{c}{Sea Surface Temperature} & \multicolumn{4}{c}{Latent Heat Flux} \\
 & &  $w=1$ & $w=2$ & $w=3$ & $w=4$ & $w=1$ & $w=2$ & $w=3$ & $w=4$ \\
 \hline
\multirow{3}{*}{$Y_1 =$ TS, $p$} 
			&$r=1$  & 0.02 - & 0.00  & \textbf{0.25} - & 0.00  & 0.00 & 0.00 & 0.00  & 0.00  \\ 
                         &$r=2$  &  \textbf{0.10} - & 0.07 - & 0.08 - & 0.00  & 0.01 - & \textcolor{red}{\textbf{0.67}} - & 0.00  & \textbf{0.44} -\\ 
                         &$r=3$  & \textbf{0.45} - & \textbf{0.22} - & 0.00 - & 0.00  & \textbf{0.21} + & 0.00 & 0.00 & 0.00 \\ 
                         &$r=4$  & \textcolor{red}{\textbf{0.48}} - & 0.01 - & 0.14 - & 0.00  & 0.00  & 0.00 & \textbf{0.28} - & 0.00 \\ 
  \hline
\multirow{3}{*}{$Y_1 =$ TS, $\lambda$} 
                         & $r=1$ & 0.00   & 0.01 +& 0.00    & 0.00    & 0.00 & 0.00 & \textbf{0.12} +& 0.00 \\ 
                         &$r=2$  & \textbf{0.77} - & \textbf{0.11} - & \textbf{0.64} -  & \textbf{0.43} - & 0.05 +& \textcolor{red}{\textbf{0.37}} - & \textbf{0.10} +& \textcolor{red}{\textbf{0.20}} -\\ 
                         &$r=3$  & \textbf{0.81} - & \textbf{0.71} - & \textbf{0.17} -  & \textbf{0.27} - & 0.01 +& \textcolor{red}{\textbf{0.40}} +& \textcolor{red}{\textbf{0.35}} -& 0.04 +\\ 
                         &$r=4$  & 0.01 - & 0.00   & 0.00    & 0.00    & \textbf{0.12} - & 0.00 & \textbf{0.36} - & 0.09 -\\ 
\hline  
\multirow{3}{*}{$Y_2 =$ HU, $p$} 
                         &$r=1$ & 0.00 & 0.00 & 0.00 & 0.00 & 0.02 - & 0.00 & \textcolor{red}{\textbf{0.16 }}+ & \textcolor{red}{\textbf{0.17}} -\\ 
                         &$r=2$ & \textbf{0.32} - & 0.00 & 0.00 & 0.00 & 0.00 & 0.00 & 0.00 & 0.00 \\ 
                         &$r=3$& 0.00 & 0.00 & 0.00 & 0.00 & 0.00 & 0.00 & 0.00 & 0.00 \\ 
                         &$r=4$ & 0.00 & 0.00 & 0.00 & 0.00 & 0.04 +& \textbf{0.18}+& \textbf{0.18} -& 0.06 -\\ 
\hline
\multirow{3}{*}{$Y_2 =$ HU, $\lambda$} 
                         & $r=1$  & \textcolor{red}{\textbf{0.25}} +& \textbf{0.23} -& \textbf{0.16 }-& 0.03 -& 0.09 -& 0.04 +& \textbf{0.20} +& \textcolor{red}{\textbf{0.21}} -\\ 
                         &$r=2$   & \textbf{0.35} - & 0.08 +& \textcolor{red}{\textbf{0.22}} -& 0.05 +& 0.01 +&  \textbf{0.14} +&  \textbf{0.11} -& \textbf{0.29} -\\ 
                         &$r=3$  & 0.07 -&  \textbf{0.13} -& 0.04 -& 0.04 -& 0.07 -& 0.04 +&  \textbf{0.11} +& \textbf{0.18} + \\ 
                         &$r=4$  & \textbf{0.33} -& \textbf{0.19} +& \textbf{0.22} +& 0.09 +& \textbf{0.15} +& \textbf{0.28} +& \textbf{0.19} -& \textcolor{red}{\textbf{0.12}} -\\ 
\hline
 \multirow{3}{*}{$Y_3 =$ MH, $p$} 
                         & $r=1$ & 0.00 & 0.00 & 0.00 & 0.00 & 0.00 & 0.00 &  \textbf{0.12} +& 0.00 \\ 
                         &$r=2$  & 0.00 & 0.00 & 0.00 & 0.00 & 0.00 & 0.00 & 0.00 & \textbf{0.32}- \\
                         &$r=3$  & 0.06 -& 0.03 -& 0.00 & 0.00 & 0.00 & 0.05- & \textcolor{red}{\textbf{0.25} }-& 0.00 \\ 
                         &$r=4$  & 0.00 & 0.00 & 0.00 & 0.00 & 0.00 & 0.00 & 0.00 & 0.00 \\ 
 \hline
\multirow{3}{*}{$Y_3 =$ MH, $\lambda$} 
                         & $r=1$  &\textcolor{red}{\textbf{0.17 }}+& 0.04 -& 0.00 & 0.04 +& 0.08 +& 0.04 +& \textbf{0.35} +& 0.03 +\\ 
                         &$r=2$ & 0.03  -&  \textcolor{red}{\textbf{0.10}} +& \textcolor{red}{\textbf{0.15}} -& 0.03 +& 0.05 -& 0.00 & 0.00 & \textcolor{red}{\textbf{0.30}} -\\ 
                         &$r=3$  & \textbf{0.17} -&  \textbf{0.10} -& 0.06 +& 0.04 +& 0.06 -& 0.04 -& \textbf{0.35} -& 0.06 -\\ 
                         &$r=4$  & 0.07 -& 0.07 +& 0.05 -& 0.11 +& 0.06 +&  \textbf{0.13} +& 0.07 -&  \textcolor{red}{\textbf{0.10}} -\\ 
                          \bottomrule
\end{tabular}
\end{adjustbox}
\label{tab:RES1}
\end{table}

\section{MSE per location and per year}\label{appF}
\begin{figure}[H]
\begin{center}
\includegraphics[width=0.9\textwidth]{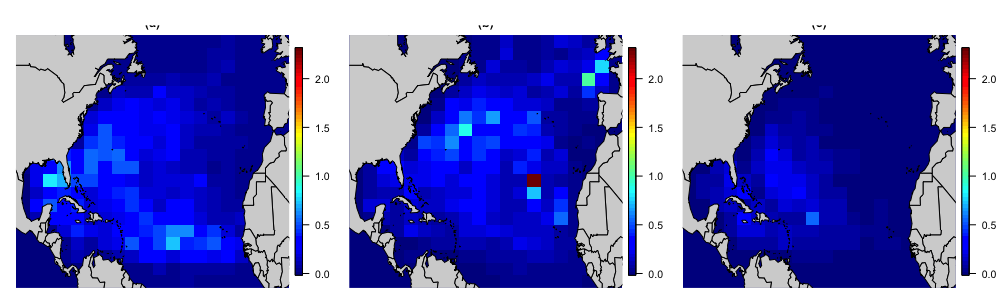}
\caption{MSE per location for each response: (a) Tropical Cyclones (TC), (b) Hurricanes (HU) and (c) Major Hurricanes (MH).}
\label{fig:MSEloc}
\end{center}
\end{figure}

\begin{figure}[H]
\begin{center}
\includegraphics[width=0.9\textwidth]{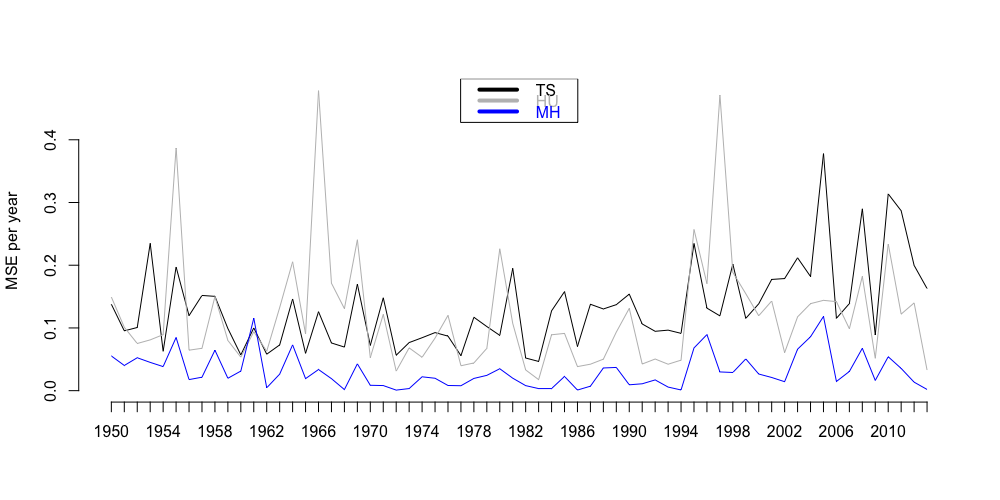}
\caption{MSE per year for each response: Tropical Cyclones (TC), Hurricanes (HU) and Major Hurricanes (MH). }
\label{fig:MSEyear}
\end{center}
\end{figure}

\section{Significant Factors for Summer and Fall}\label{appG}

\begin{figure}[H]
\begin{center}
\includegraphics[width=0.7\textwidth]{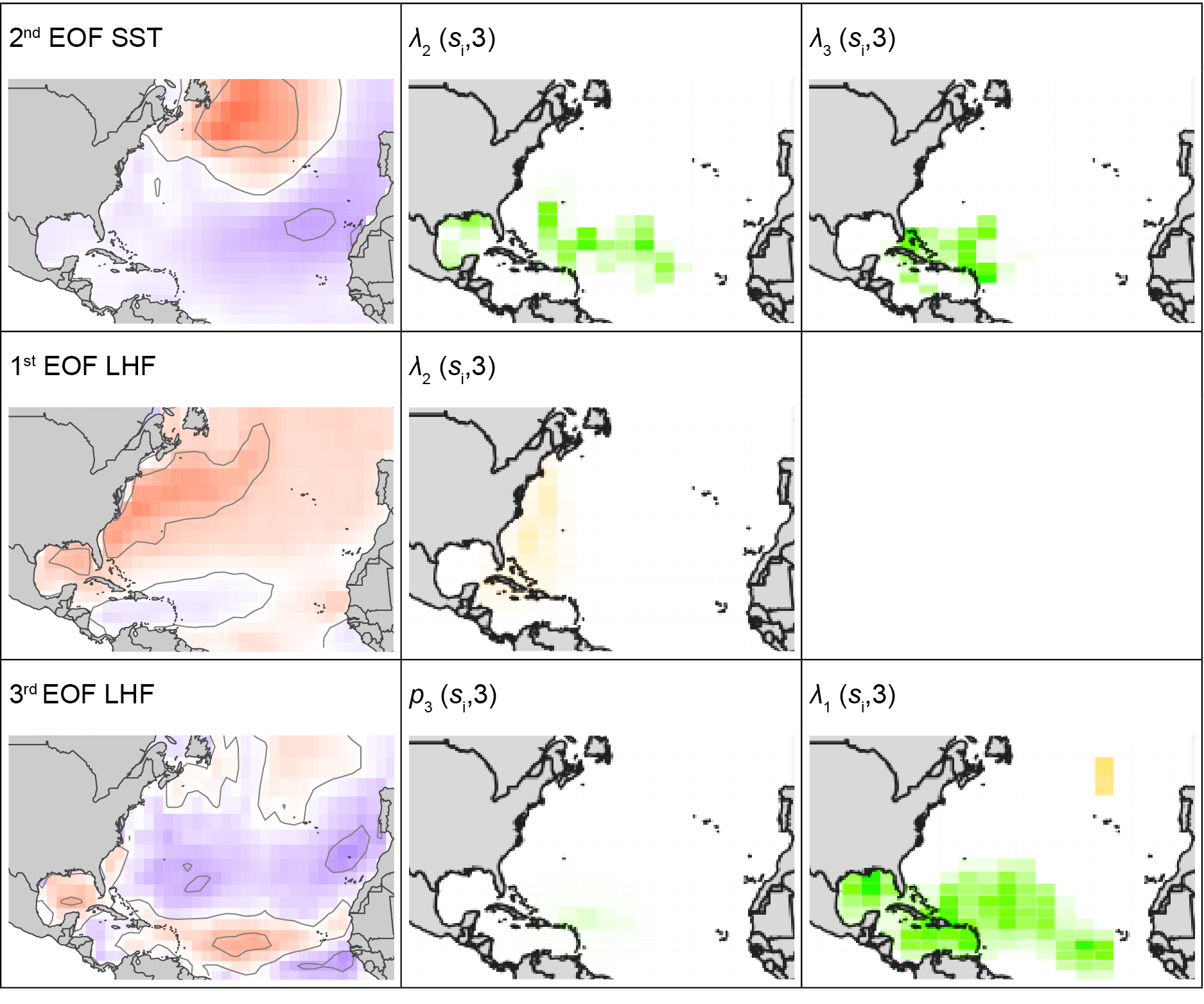}
\caption{Summer: SST or LHF Features (first left column) and their respective statistically significant associated factors. Warm colors (red, orange) are positive, and cold colors (purple, green) are negative. EOFs range from -0.2 to 0.2 and coefficients from -1 to 1.}
\label{fig:summer}
\end{center}
\end{figure}

\begin{figure}[H]
\begin{center}
\includegraphics[width=0.7\textwidth]{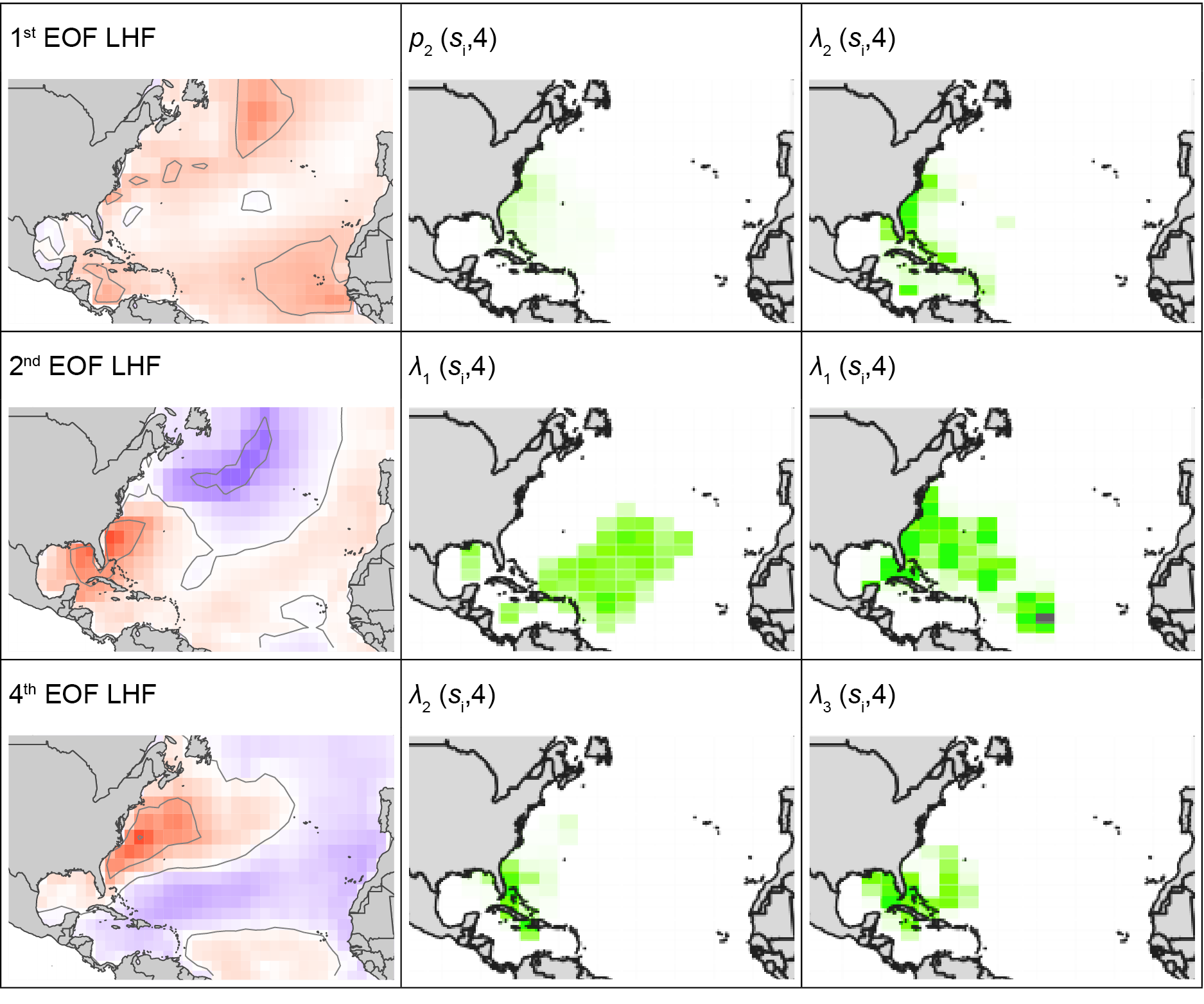}
\caption{Fall: LHF Features (first left column) and their respective statistically significant associated factors. Warm colors (red, orange) are positive, and cold colors (purple, green) are negative. EOFs range from -0.2 to 0.2 and coefficients from -1 to 1.}
\label{fig:fall}
\end{center}
\end{figure}


\section*{Acknowledgements}
We are grateful for the support provided by the National Science Foundation's Research Network for Statistical Methods for Atmospheric and Oceanic Sciences, award DMS-1107046, and National Science Foundation Awards 1406016 and 1613219. Marcela Alfaro C\'{o}rdoba is also grateful for the support provided by CONICIT Costa Rica and the University of Costa Rica. 

\begin{supplement}
\sname{Supplement A}\label{suppA}
\stitle{Code for the MCMC sampler}
\slink[url]{.zip file}
\sdescription{Code for the MCMC sampler used in the Simulation.}
\end{supplement}

\begin{supplement}
\sname{Supplement B}\label{suppB}
\stitle{Supplementary Plots}
\slink[url]{pdf file}
\sdescription{We present supplementary plots to the article. They include an example of the simulation settings realizations for covariate 1, a plot of time series for each $\xi_{\ell, r}(t,w)$, a complete plot of statistically significant factors and a summary of significant factors for summer and fall.}
\end{supplement}

\bibliographystyle{imsart}
\bibliography{Methods1_v4}

\begin{thebibliography}{26}

\bibitem[\protect\citeauthoryear{Blake and Gray}{2004}]{Blake2004}
\begin{barticle}[author]
\bauthor{\bsnm{Blake},~\bfnm{Eric~S.}\binits{E.~S.}} \AND
  \bauthor{\bsnm{Gray},~\bfnm{William~M.}\binits{W.~M.}}
(\byear{2004}).
\btitle{{Prediction of August Atlantic Basin Hurricane Activity}}.
\bjournal{Weather and Forecasting}
\bvolume{19}
\bpages{1044--1060}.
\bdoi{10.1175/814.1}
\end{barticle}
\endbibitem

\bibitem[\protect\citeauthoryear{{Boehm Vock} et~al.}{2015}]{BoehmVock2015}
\begin{barticle}[author]
\bauthor{\bsnm{{Boehm Vock}},~\bfnm{Laura~F}\binits{L.~F.}},
  \bauthor{\bsnm{Reich},~\bfnm{Brian~J}\binits{B.~J.}},
  \bauthor{\bsnm{Fuentes},~\bfnm{Montserrat}\binits{M.}} \AND
  \bauthor{\bsnm{Dominici},~\bfnm{Francesca}\binits{F.}}
(\byear{2015}).
\btitle{{Spatial variable selection methods for investigating acute health
  effects of fine particulate matter components.}}
\bjournal{Biometrics}
\bvolume{71}
\bpages{167--77}.
\bdoi{10.1111/biom.12254}
\end{barticle}
\endbibitem

\bibitem[\protect\citeauthoryear{Cai et~al.}{2013}]{Cai2013}
\begin{barticle}[author]
\bauthor{\bsnm{Cai},~\bfnm{Bo}\binits{B.}},
  \bauthor{\bsnm{Lawson},~\bfnm{Andrew~B}\binits{A.~B.}},
  \bauthor{\bsnm{Hossain},~\bfnm{Monir}\binits{M.}},
  \bauthor{\bsnm{Choi},~\bfnm{Jungsoon}\binits{J.}},
  \bauthor{\bsnm{Kirby},~\bfnm{Russell~S}\binits{R.~S.}} \AND
  \bauthor{\bsnm{Liu},~\bfnm{Jihong}\binits{J.}}
(\byear{2013}).
\btitle{{Bayesian semiparametric model with spatially-temporally varying
  coefficients selection.}}
\bjournal{Statistics in medicine}
\bvolume{32}
\bpages{3670--85}.
\bdoi{10.1002/sim.5789}
\end{barticle}
\endbibitem

\bibitem[\protect\citeauthoryear{Chipman}{1996}]{Chipman1996}
\begin{barticle}[author]
\bauthor{\bsnm{Chipman},~\bfnm{Hugh}\binits{H.}}
(\byear{1996}).
\btitle{{Bayesian variable selection with related predictors}}.
\bjournal{Canadian Journal of Statistics}
\bvolume{24}
\bpages{17--36}.
\bdoi{10.2307/3315687}
\end{barticle}
\endbibitem

\bibitem[\protect\citeauthoryear{Elsner and Jagger}{2006}]{Elsner2006}
\begin{barticle}[author]
\bauthor{\bsnm{Elsner},~\bfnm{James~B.}\binits{J.~B.}} \AND
  \bauthor{\bsnm{Jagger},~\bfnm{Thomas~H.}\binits{T.~H.}}
(\byear{2006}).
\btitle{{Prediction Models for Annual U.S. Hurricane Counts}}.
\bjournal{Journal of Climate}
\bvolume{19}
\bpages{2935--2952}.
\bdoi{10.1175/JCLI3729.1}
\end{barticle}
\endbibitem

\bibitem[\protect\citeauthoryear{Elsner, Liu and Kocher}{2000}]{Elsner2000}
\begin{barticle}[author]
\bauthor{\bsnm{Elsner},~\bfnm{James~B.}\binits{J.~B.}},
  \bauthor{\bsnm{Liu},~\bfnm{Kam-biu}\binits{K.-b.}} \AND
  \bauthor{\bsnm{Kocher},~\bfnm{Bethany}\binits{B.}}
(\byear{2000}).
\btitle{{Spatial Variations in Major U.S. Hurricane Activity: Statistics and a
  Physical Mechanism}}.
\bjournal{Journal of Climate}
\bvolume{13}
\bpages{2293--2305}.
\bdoi{10.1175/1520-0442(2000)013<2293:SVIMUS>2.0.CO;2}
\end{barticle}
\endbibitem

\bibitem[\protect\citeauthoryear{Gelfand et~al.}{2003}]{Gelfand2003a}
\begin{barticle}[author]
\bauthor{\bsnm{Gelfand},~\bfnm{Alan~E}\binits{A.~E.}},
  \bauthor{\bsnm{Kim},~\bfnm{Hyon-Jung}\binits{H.-J.}},
  \bauthor{\bsnm{Sirmans},~\bfnm{C.~F}\binits{C.~F.}} \AND
  \bauthor{\bsnm{Banerjee},~\bfnm{Sudipto}\binits{S.}}
(\byear{2003}).
\btitle{{Spatial Modeling With Spatially Varying Coefficient Processes}}.
\bjournal{Journal of the American Statistical Association}
\bvolume{98}
\bpages{387--396}.
\bdoi{10.1198/016214503000170}
\end{barticle}
\endbibitem

\bibitem[\protect\citeauthoryear{George and McCulloch}{1993}]{George1993}
\begin{barticle}[author]
\bauthor{\bsnm{George},~\bfnm{Edward~I.}\binits{E.~I.}} \AND
  \bauthor{\bsnm{McCulloch},~\bfnm{Robert~E.}\binits{R.~E.}}
(\byear{1993}).
\btitle{{Variable Selection via Gibbs Sampling}}.
\bjournal{Journal of the American Statistical Association}
\bvolume{88}
\bpages{881--889}.
\bdoi{10.1080/01621459.1993.10476353}
\end{barticle}
\endbibitem

\bibitem[\protect\citeauthoryear{Gray et~al.}{1994}]{Gray1994}
\begin{barticle}[author]
\bauthor{\bsnm{Gray},~\bfnm{William~M.}\binits{W.~M.}},
  \bauthor{\bsnm{Landsea},~\bfnm{Christopher~W.}\binits{C.~W.}},
  \bauthor{\bsnm{Mielke},~\bfnm{Paul~W.}\binits{P.~W.}} \AND
  \bauthor{\bsnm{Berry},~\bfnm{Kenneth~J.}\binits{K.~J.}}
(\byear{1994}).
\btitle{{Predicting Atlantic Basin Seasonal Tropical Cyclone Activity by 1
  June}}.
\bjournal{Weather and Forecasting}
\bvolume{9}
\bpages{103--115}.
\bdoi{10.1175/1520-0434(1994)009<0103:PABSTC>2.0.CO;2}
\end{barticle}
\endbibitem

\bibitem[\protect\citeauthoryear{Hodges, Jagger and Elsner}{2014}]{Hodges2014}
\begin{barticle}[author]
\bauthor{\bsnm{Hodges},~\bfnm{Robert~E.}\binits{R.~E.}},
  \bauthor{\bsnm{Jagger},~\bfnm{Thomas~H.}\binits{T.~H.}} \AND
  \bauthor{\bsnm{Elsner},~\bfnm{James~B.}\binits{J.~B.}}
(\byear{2014}).
\btitle{{The sun-hurricane connection: Diagnosing the solar impacts on
  hurricane frequency over the North Atlantic basin using a space–time
  model}}.
\bjournal{Natural Hazards}.
\bdoi{10.1007/s11069-014-1120-9}
\end{barticle}
\endbibitem

\bibitem[\protect\citeauthoryear{Ishwaran and Rao}{2005}]{Ishwaran2005}
\begin{barticle}[author]
\bauthor{\bsnm{Ishwaran},~\bfnm{Hemant}\binits{H.}} \AND
  \bauthor{\bsnm{Rao},~\bfnm{J.~Sunil}\binits{J.~S.}}
(\byear{2005}).
\btitle{{Spike and slab variable selection: Frequentist and Bayesian
  strategies}}.
\bjournal{The Annals of Statistics}
\bvolume{33}
\bpages{730--773}.
\bdoi{10.1214/009053604000001147}
\end{barticle}
\endbibitem

\bibitem[\protect\citeauthoryear{Jagger, Niu and Elsner}{2002}]{Jagger2002}
\begin{barticle}[author]
\bauthor{\bsnm{Jagger},~\bfnm{Thomas~H.}\binits{T.~H.}},
  \bauthor{\bsnm{Niu},~\bfnm{Xufeng}\binits{X.}} \AND
  \bauthor{\bsnm{Elsner},~\bfnm{James~B.}\binits{J.~B.}}
(\byear{2002}).
\btitle{{A space-time model for seasonal hurricane prediction}}.
\bjournal{International Journal of Climatology}
\bvolume{22}
\bpages{451--465}.
\bdoi{10.1002/joc.755}
\end{barticle}
\endbibitem

\bibitem[\protect\citeauthoryear{Kalnay et~al.}{1996}]{Kalnay1996}
\begin{barticle}[author]
\bauthor{\bsnm{Kalnay},~\bfnm{E.}\binits{E.}},
  \bauthor{\bsnm{Kanamitsu},~\bfnm{M.}\binits{M.}},
  \bauthor{\bsnm{Kistler},~\bfnm{R.}\binits{R.}},
  \bauthor{\bsnm{Collins},~\bfnm{W.}\binits{W.}},
  \bauthor{\bsnm{Deaven},~\bfnm{D.}\binits{D.}},
  \bauthor{\bsnm{Gandin},~\bfnm{L.}\binits{L.}},
  \bauthor{\bsnm{Iredell},~\bfnm{M.}\binits{M.}},
  \bauthor{\bsnm{Saha},~\bfnm{S.}\binits{S.}},
  \bauthor{\bsnm{White},~\bfnm{G.}\binits{G.}},
  \bauthor{\bsnm{Woollen},~\bfnm{J.}\binits{J.}},
  \bauthor{\bsnm{Zhu},~\bfnm{Y.}\binits{Y.}},
  \bauthor{\bsnm{Leetmaa},~\bfnm{A.}\binits{A.}},
  \bauthor{\bsnm{Reynolds},~\bfnm{R.}\binits{R.}},
  \bauthor{\bsnm{Chelliah},~\bfnm{M.}\binits{M.}},
  \bauthor{\bsnm{Ebisuzaki},~\bfnm{W.}\binits{W.}},
  \bauthor{\bsnm{Higgins},~\bfnm{W.}\binits{W.}},
  \bauthor{\bsnm{Sepowiak},~\bfnm{J.}\binits{J.}},
  \bauthor{\bsnm{Mo},~\bfnm{K.~C.}\binits{K.~C.}},
  \bauthor{\bsnm{Ropelewski},~\bfnm{C.}\binits{C.}},
  \bauthor{\bsnm{Wang},~\bfnm{J.}\binits{J.}},
  \bauthor{\bsnm{Jenne},~\bfnm{Roy}\binits{R.}} \AND
  \bauthor{\bsnm{Joseph},~\bfnm{Dennis}\binits{D.}}
(\byear{1996}).
\btitle{{The NCEP/NCAR 40-Year Reanalysis Project}}.
\bjournal{Bulletin of the American Meteorological Society}
\bvolume{77}
\bpages{437--471}.
\bdoi{10.1175/1520-0477(1996)077<0437:TNYRP>2.0.CO;2}
\end{barticle}
\endbibitem

\bibitem[\protect\citeauthoryear{Keith and Xie}{2009}]{Keith2009}
\begin{barticle}[author]
\bauthor{\bsnm{Keith},~\bfnm{Elinor}\binits{E.}} \AND
  \bauthor{\bsnm{Xie},~\bfnm{Lian}\binits{L.}}
(\byear{2009}).
\btitle{{Predicting Atlantic Tropical Cyclone Seasonal Activity in April}}.
\bjournal{Weather and Forecasting}
\bvolume{24}
\bpages{436--455}.
\bdoi{10.1175/2008WAF2222139.1}
\end{barticle}
\endbibitem

\bibitem[\protect\citeauthoryear{Lehmiller, Kimberlain and
  Elsner}{1997}]{Lehmiller1997}
\begin{barticle}[author]
\bauthor{\bsnm{Lehmiller},~\bfnm{G.~S.}\binits{G.~S.}},
  \bauthor{\bsnm{Kimberlain},~\bfnm{T.~B.}\binits{T.~B.}} \AND
  \bauthor{\bsnm{Elsner},~\bfnm{J.~B.}\binits{J.~B.}}
(\byear{1997}).
\btitle{{Seasonal Prediction Models for North Atlantic Basin Hurricane
  Location}}.
\bjournal{Monthly Weather Review}
\bvolume{125}
\bpages{1780--1791}.
\bdoi{10.1175/1520-0493(1997)125<1780:SPMFNA>2.0.CO;2}
\end{barticle}
\endbibitem

\bibitem[\protect\citeauthoryear{Lorenz}{1956}]{Lorenz1956}
\begin{barticle}[author]
\bauthor{\bsnm{Lorenz},~\bfnm{E~N}\binits{E.~N.}}
(\byear{1956}).
\btitle{{Empirical Orthogonal Functions and Statistical Weather Prediction}}.
\bjournal{Technical report Statistical Forecast Project Report 1 Department of
  Meteorology MIT 49}
\bvolume{1}
\bpages{52}.
\end{barticle}
\endbibitem

\bibitem[\protect\citeauthoryear{Lum}{2012}]{Lum2012}
\begin{barticle}[author]
\bauthor{\bsnm{Lum},~\bfnm{Kristian}\binits{K.}}
(\byear{2012}).
\btitle{{Bayesian variable selection for spatially dependent generalized linear
  models}}.
\end{barticle}
\endbibitem

\bibitem[\protect\citeauthoryear{Neelon, Ghosh and Loebs}{2013}]{Neelon2013}
\begin{barticle}[author]
\bauthor{\bsnm{Neelon},~\bfnm{Brian}\binits{B.}},
  \bauthor{\bsnm{Ghosh},~\bfnm{Pulak}\binits{P.}} \AND
  \bauthor{\bsnm{Loebs},~\bfnm{Patrick~F}\binits{P.~F.}}
(\byear{2013}).
\btitle{{A Spatial Poisson Hurdle Model for Exploring Geographic Variation in
  Emergency Department Visits.}}
\bjournal{Journal of the Royal Statistical Society. Series A, (Statistics in
  Society)}
\bvolume{176}
\bpages{389--413}.
\bdoi{10.1111/j.1467-985X.2012.01039.x}
\end{barticle}
\endbibitem

\bibitem[\protect\citeauthoryear{NHC}{2016}]{NHC2016}
\begin{bmisc}[author]
\bauthor{\bsnm{NHC}}
(\byear{2016}).
\btitle{{Eye, Eyewall and Tropical Cyclone. Glossary of National Hurricane
  Center Terms}}.
\end{bmisc}
\endbibitem

\bibitem[\protect\citeauthoryear{Reich et~al.}{2010}]{Reich2010}
\begin{barticle}[author]
\bauthor{\bsnm{Reich},~\bfnm{Brian~J}\binits{B.~J.}},
  \bauthor{\bsnm{Fuentes},~\bfnm{Montserrat}\binits{M.}},
  \bauthor{\bsnm{Herring},~\bfnm{Amy~H}\binits{A.~H.}} \AND
  \bauthor{\bsnm{Evenson},~\bfnm{Kelly~R}\binits{K.~R.}}
(\byear{2010}).
\btitle{{Bayesian variable selection for multivariate spatially varying
  coefficient regression.}}
\bjournal{Biometrics}
\bvolume{66}
\bpages{772--82}.
\bdoi{10.1111/j.1541-0420.2009.01333.x}
\end{barticle}
\endbibitem

\bibitem[\protect\citeauthoryear{Scheel et~al.}{2013}]{Scheel2013}
\begin{barticle}[author]
\bauthor{\bsnm{Scheel},~\bfnm{Ida}\binits{I.}},
  \bauthor{\bsnm{Ferkingstad},~\bfnm{Egil}\binits{E.}},
  \bauthor{\bsnm{Frigessi},~\bfnm{Arnoldo}\binits{A.}},
  \bauthor{\bsnm{Haug},~\bfnm{Ola}\binits{O.}},
  \bauthor{\bsnm{Hinnerichsen},~\bfnm{Mikkel}\binits{M.}} \AND
  \bauthor{\bsnm{Meze-Hausken},~\bfnm{Elisabeth}\binits{E.}}
(\byear{2013}).
\btitle{{A Bayesian hierarchical model with spatial variable selection: the
  effect of weather on insurance claims}}.
\bjournal{Journal of the Royal Statistical Society: Series C (Applied
  Statistics)}
\bvolume{62}
\bpages{85--100}.
\bdoi{10.1111/j.1467-9876.2012.01039.x}
\end{barticle}
\endbibitem

\bibitem[\protect\citeauthoryear{Smith and Fahrmeir}{2007}]{Smith2007}
\begin{barticle}[author]
\bauthor{\bsnm{Smith},~\bfnm{Michael}\binits{M.}} \AND
  \bauthor{\bsnm{Fahrmeir},~\bfnm{Ludwig}\binits{L.}}
(\byear{2007}).
\btitle{{Spatial Bayesian Variable Selection With Application to Functional
  Magnetic Resonance Imaging}}.
\bjournal{Journal of the American Statistical Association}
\bvolume{102}
\bpages{417--431}.
\bdoi{10.1198/016214506000001031}
\end{barticle}
\endbibitem

\bibitem[\protect\citeauthoryear{Werner and Holbrook}{2011}]{Werner2011}
\begin{barticle}[author]
\bauthor{\bsnm{Werner},~\bfnm{Angelika}\binits{A.}} \AND
  \bauthor{\bsnm{Holbrook},~\bfnm{Neil~J.}\binits{N.~J.}}
(\byear{2011}).
\btitle{{A Bayesian Forecast Model of Australian Region Tropical Cyclone
  Formation}}.
\bjournal{Journal of Climate}
\bvolume{24}
\bpages{6114--6131}.
\bdoi{10.1175/2011JCLI4231.1}
\end{barticle}
\endbibitem

\bibitem[\protect\citeauthoryear{Xie}{2005}]{Xie2005}
\begin{barticle}[author]
\bauthor{\bsnm{Xie},~\bfnm{Lian}\binits{L.}}
(\byear{2005}).
\btitle{{The effect of Atlantic sea surface temperature dipole mode on
  hurricanes: Implications for the 2004 Atlantic hurricane season}}.
\bjournal{Geophysical Research Letters}
\bvolume{32}
\bpages{L03701}.
\bdoi{10.1029/2004GL021702}
\end{barticle}
\endbibitem

\bibitem[\protect\citeauthoryear{Xie and Liu}{2014}]{Xie2014b}
\begin{barticle}[author]
\bauthor{\bsnm{Xie},~\bfnm{Kenny}\binits{K.}} \AND
  \bauthor{\bsnm{Liu},~\bfnm{Bin}\binits{B.}}
(\byear{2014}).
\btitle{{An ENSO-Forecast Independent Statistical Model for the Prediction of
  Annual Atlantic Tropical Cyclone Frequency in April}}.
\bjournal{Advances in Meteorology}
\bvolume{2014}
\bpages{1--11}.
\bdoi{10.1155/2014/248148}
\end{barticle}
\endbibitem

\bibitem[\protect\citeauthoryear{Xie et~al.}{2014}]{Xie2014}
\begin{btechreport}[author]
\bauthor{\bsnm{Xie},~\bfnm{Lian}\binits{L.}},
  \bauthor{\bsnm{Alfaro-C\'{o}rdoba},~\bfnm{Marcela}\binits{M.}},
  \bauthor{\bsnm{Liu},~\bfnm{Bin}\binits{B.}} \AND
  \bauthor{\bsnm{Fuentes},~\bfnm{Montserrat}\binits{M.}}
(\byear{2014}).
\btitle{{2014 Atlantic Tropical Cyclone Outlook }}
\btype{Technical Report},
\bpublisher{North Carolina State University},
\baddress{Raleigh}.
\end{btechreport}
\endbibitem

\end{thebibliography}

\end{document}